\documentclass[fleqn,usenatbib]{mnras}

\usepackage[T1]{fontenc}

\DeclareRobustCommand{\VAN}[3]{#2}
\let\VANthebibliography\thebibliography
\def\thebibliography{\DeclareRobustCommand{\VAN}[3]{##3}\VANthebibliography}

\usepackage{graphicx}	
\usepackage{amsmath}	
\usepackage{amssymb}	
\usepackage{newtxtext,newtxmath}

\title[Passive galaxies in modified gravity]{Galaxy evolution in modified gravity simulations: using passive galaxies to constrain gravity with upcoming surveys}

\author[Diego Pallero et al.]{
Diego Pallero,$^{1}$\thanks{E-mail: diego.pallero@usm.cl},
Facundo A. G\'omez$^{2,3}$,
Nelson D. Padilla$^{4}$,
Y. L. Jaff\'e$^{1,5}$,
Carlton M. Baugh$^{6}$,\newauthor
Baojiu Li$^{6}$,
C\'esar Hern\'andez-Aguayo$^{7,8}$,
Christian Arnold$^{6}$
\\
$^{1}$Departamento de F\'isica, Universidad T\'ecnica Federico Santa Mar\'ia, Avenida Espa\~na 1600, Valpara\'iso, Chile\\ 
$^{2}$Departamento de astronom\'ia, Universidad de La Serena, Avenida Juan Cisternas 1200, La Serena, Chile\\
$^{3}$Instituto de Investigaci\'on Multidisciplinar en Ciencia y Tecnolog\'ia, Universidad de La Serena, Ra\'ul Bitr\'an 1305, La Serena, Chile\\
$^{4}$Instituto de Astrom\'ia Te\'orica y Experimental (IATE), CONICET Universidad Nacional de C\'ordoba, Laprida 854, X5000 BGR, C\'ordoba, Argentina\\
$^{5}$Instituto de F\'isica y Astronom\'ia, Universidad de Valpara\'iso, 1111 Gran Breta\~na, Valpara\'iso, Chile\\
$^{6}$Institute for Computational Cosmology, Department of Physics, Durham University, SouthRoad, Durham, DH13LE, UK\\
$^{7}$Max-Planck-Institutf{\"u}r Astrophysik, Karl-Schwarzschild-Str. 1, D-85748, Garching, Germany\\
$^{8}$Excellence Cluster ORIGINS, Boltzmannstrasse 2, D-85748 Garching, Germany
}

\date{Accepted XXX. Received YYY; in original form ZZZ}

\pubyear{2015}

\begin{document}
\label{firstpage}
\pagerange{\pageref{firstpage}--\pageref{lastpage}}
\maketitle

\begin{abstract}
We present a quantitative analysis of the properties of galaxies and structures evolving in universes dominated by different modified gravitational models, including two variants of the f(R)-gravity (F) and two of the Dvali-Gabdadze-Poratti (N) braneworld model, which respectively feature the chameleon and Vainshtein screening mechanisms. Using the Simulation HYdrodynamics BeyONd Einstein (SHYBONE) cosmological hydrodynamical full-physics simulations suite, we study the departures in the properties of galaxies residing in different environments with respect to the standard model (GR). Using two different criteria to compare, we find that structures formed within modified gravity tend to show a denser gas density profile than their GR counterparts. Within the different modified gravity models, N1 and F5 gravity models show greater departures from the standard model, with gas density profiles $\rho_{\rm IGM} \geq 30\%$ denser in the outskirts for the N1 model, and in the inner parts for the F5 model. 
Additionally, we find that haloes evolving in MG universes show, in general, larger quenched fractions than GR, reaching up to $20\%$ larger quenching fractions in F5 regardless of the stellar mass of the galaxy.
With respect to the other models, F6, N1 and N5 show slightly larger quenched fractions, but no strong differences can be found.
These results directly impact the colour distribution of galaxies, making them in MG models redder and older than their GR counterparts.
Like GR, once the environment starts to play a role, galaxies rapidly get quenched and the differences between models vanish.
\end{abstract}




\section{Introduction}
To understand how the Universe evolves, it is necessary to characterize its composition and how its different constituents interact with each other.
All matter in the Universe interacts gravitationally, but a deep understanding of the physical nature of gravity has proven to be  hard to achieve. The current standard cosmological model ($\Lambda$CDM) is constructed with Einstein's General Relativity (GR) as its gravitational model and has as its main constituents today the cosmological constant $\Lambda$ and the cold dark matter. The $\Lambda$CDM model has provided a simple yet very accurate description of the Universe \citep[eg.][]{BOSS, Planck16,Planck18}. Many observational pieces of evidence support it. However, even though GR has been empirically confirmed on small scales, the lack of accuracy tests at cosmological scales has allowed constraining this model of gravity only to a small degree at large scales.

With the arrival of new galaxy surveys such as the Dark Energy Spectroscopic Instrument \citep[DESI; ][]{DESI}, EUCLID \citep{EUCLID},  and the Vera C. Rubin Observatory, 
\citep[LSST; ][]{LSST}, gravity will be testable to an unprecedented level of accuracy at all scales. This will allow astronomers to distinguish between different gravitational models. However, in order to do so, it is necessary to fully understand the impact that different models could have on the distribution of galaxies in the large-scale structure, and the imprints that they could leave in their properties.

Within this context, hydrodynamical simulations play a key role in connecting theoretical predictions with observations. 
Even though there have been some attempts to follow galaxy formation in alternative gravity models using semi-analytics \citep[eg.][]{Fontanot13, Fontanot15a, Fontaton15b}, these calculations have not tended to include as many effects that are changed by modified gravity as in hydrodynamical simulations.

In the last decade, hydrodynamical simulations of large cosmological volumes \citep[eg.][]{Vogelsberger14, Schaye15, Pillepich18a, Nelson18,Schaye23,Pakmor23}
have successfully reproduced observable properties of single galaxies as well as populations of galaxies. Simulations have also been widely used as a tool to study the evolution of galaxies through cosmic time, providing accurate descriptions of some complex phenomena experienced throughout their history. Nevertheless, the lack of models using non-standard cosmologies hinders the comparison between observations and modified gravity (MG) theories. The first steps have been given in an effort to study the impact that some of the most studied models of MG could have in the evolution of galaxies \citep{Llinares14,Hammami15,Arnold14,Arnold15,Arnold16,Ellewsen18}. 

However, these early works usually do not have a cosmological volume big enough to make statistically representative studies or adopt a \textit{`full physics'} model that allows a detailed description of the evolution of baryons within these models. Moreover, these early works usually have a limited mass range of haloes studied and lack a wide variety of different environments.

In this work, we use a more recent set of simulations from the Simulation HYdrodynamics BeyONd Einstein project, \textsc{shybone}. The project introduces the first suite of cosmological simulations with a comprehensive galaxy formation model, evolved with two of the most studied modified gravity models, the Hu \& Sawicki $f(R)-$gravity \citep{Hu07},  and the normal Dvali-Gabadadze-Porrati (nDGP) \citep{Dvali00} brane model in the Newtonian limit. The simulations use the same galaxy formation model employed in the Illustris-TNG simulation \citep{Pillepich18a}. The suite is made up of several runs, from small high-resolution  (25cMpc) to big intermediate-resolution boxes (100cMpc). With these state-of-the-art simulations, it is now possible to study galaxy evolution beyond the classical standard model. Moreover, thanks to the detailed description provided by the galaxy formation model of Illustris-TNG simulations, it is possible to understand how different gravitational models could affect key galaxy properties as we observe them nowadays.

In our current cosmological model, it is well known that the environment in which a galaxy resides plays a decisive role in shaping its properties \citep{Dressler80,Dressler84,Poggianti01}. In particular, \citet{Dressler80} showed that galaxy clusters possess an excess of early-type galaxies in comparison to the field. Moreover, galaxies residing in these high-density environments tends to show redder colours than galaxies in the field with the same stellar mass, as reported by several authors \citep{Gomez03,Kauffmann04, Poggianti06}. These colour/morphological transformations are caused by a decrease in their star formation rate, a product of the depletion of their gas content. Nevertheless, the dominant process that led to this “quenched state” is still an unanswered question. 
Interactions between galaxies inside clusters \citep{Toomre72, Moore96}, between galaxies and the intracluster medium (ICM) \citep{Gunn72,Abadi99,Jaffe15} and tidal forces produced by the potential well of the cluster \citep{Miller86, Boselli06} can all produce a decrease in the gas content and change their morphology. 
Moreover, it has been shown that there is an interplay between external and internal mechanisms \citep[eg.][]{Peng10}. As all these mechanisms are directly or indirectly related to gravity, by changing the gravitational model, the way in which galaxies may be affected can completely change, by enhancing or diminishing the discrepancies between environments.      

Here, we discuss our efforts to use the \textsc{shybone} simulations to statistically characterize some of the most studied properties of galaxies in a standard model Universe, such as passive fractions and colour distributions, as a function of galaxy stellar mass and environment. 

We will use populations of galaxies in different environments to characterize the departures between MG cosmological models with respect to the standard model. This represents one of the first attempts to characterize the properties of galaxies in different cosmological contexts based on some of the most promising candidates for modified gravity. Any clear departures between models will be readily tested thanks to available and upcoming large galaxy surveys \citep[eg.][]{DESI,EUCLID_redshift}.

This paper is organized as follows. In Section \ref{sec:gfm} we review the galaxy formation model and the properties of the simulations used in this project. Also, we review some of the key aspects of the considered modified gravity models. In Section \ref{sec:comp_haloes} we define the criteria used to compare haloes between different gravitational models. These are based on properties such as the stellar mass of central galaxies and the measured $M_{200}$ of a given halo. In Section \ref{sec:res} we show the differences in the galaxy properties between different models as a function of the environment in which they reside. Finally, in Section \ref{sec:summ} we summarize our results and discuss the next steps.

\section{Galaxy formation in alternative gravity models}
\label{sec:gfm}

Here, we introduce the \textsc{shybone}  simulation suite, a series of hydrodynamical cosmological simulations carried out with the \textsc{arepo} code \citep{Springel10} augmented with a modified gravity solver, first presented in \citep{Arnold19}.
The simulation suite is currently composed of two sets of simulations dedicated to studying different models of modified gravity. The first suite, presented in \citep{Arnold19}, has a model universe where gravity is described by the $f(R)$-gravity model \citep{Hu07}. A second simulation suite was later performed to study a universe evolved over a normal Dvali-Gabadadze-Porrati (nDGP) braneworld model \citep{Dvali00}. These simulations follow the exact simulation specifications, cosmological parameters and baryonic physics model as what was presented in \citet{Arnold19}, and were first introduced in \citet{Hernandez21}.

In what follows, we discuss the main features of the gravitational models considered, as well as present the details of the simulations.
Both simulation suites were performed including the galaxy formation model used in the Illustris-TNG simulation \citep{Springel18,Pillepich18,Dylan18,Marinacci18,Naiman18}, following the same subgrid physics prescriptions and using the same parameter choices \citep[for details about the agreement between SHYBONE and TNG see][]{Arnold19}.

\subsection{Modified Gravity Models}

\subsubsection{F(R)-gravity}

The \textit{f(R)}-gravity model is an extended version of Einstein's General Relativity, which includes an additional scalar degree of freedom \citep{Buchdahl70}. This parameter produces a so-called fifth force that yields an enhancement of gravity in low-density environments by $4/3$. Regions within deep gravitational potentials experience a chameleonic screening such that the forces experienced within them are the same as expected for GR.

To construct this model some modifications are applied to the Einstein-Hilbert action, $S$, by adding a function of the Ricci scalar curvature $R$, \textit{f(R)}, as follows:

\begin{equation}
    S = \int {\rm d}^4x\sqrt{-g}\left[\frac{R + f(R)}{16\pi G} + \mathcal{L}_{\rm M}       \right],
\end{equation}
where g is the determinant of the metric tensor $g_{\mu \nu }$, $G$ is the universal gravitational constant, and $\mathcal{L}_{\rm M}$ is the Lagrangian of the density field. With this modification, an extra tensor, $\chi_{\mu \nu}$ is added to Einstein's field equations:

\begin{equation}
    \chi_{\mu \nu} = f_R R_{\mu \nu} - \left(  \frac{f}{2} - \square f_R \right) g_{\mu \nu} - \nabla_\mu \nabla_\nu f_R. 
\end{equation}

This yields the field equations of $f(R)$-gravity model in the form  

\begin{equation}
     G_{\mu \nu} + \chi_{\mu \nu} = 8\pi GT_{\mu \nu},
\end{equation}

where, $G_{\mu \nu}$, $R_{\mu \nu}$ and $T_{\mu \nu}$ correspond, respectively, to the Einstein tensor, the Ricci tensor and the stress-energy tensor.  $\nabla _\mu$ corresponds to the covariant derivative associated with the metric tensor, and $\square$ corresponds to the d'Alembert operator, where $\square \equiv \nabla _\mu \nabla^{\nu}$.
The extra scalar degree of freedom, $f_R$, corresponds to the derivative of the scalar function $f_R \equiv df(R)/dR$ and mediates the previously mentioned `fifth force’, an attractive force exerted over massive particles. 

\begin{table*}
\resizebox{\textwidth}{!}{
\begin{tabular}{llllllll}
\hline
Simulation            & Hydro model           & Cosmologies              & $L_{\rm box}$                         & $N_{\rm DM}$          & $N_{\rm gas}$         & m$_{\rm DM}$                                & $\bar{m}_{\rm gas}$                         \\
\multicolumn{1}{c}{-} & \multicolumn{1}{c}{-} & \multicolumn{1}{c}{-}    & \multicolumn{1}{c}{{[}$h^{-1}$Mpc{]}} & \multicolumn{1}{c}{-} & \multicolumn{1}{c}{-} & \multicolumn{1}{c}{{[}$h^{-1}$M$_\odot${]}} & \multicolumn{1}{c}{{[}$h^{-1}$M$_\odot${]}} \\ \hline
Full-physics, L62     & TNG-model             & $\Lambda$CDM, F6, F5     & 62                                    & 512$^3$               & 512$^3$               & 1.3$\times 10^8$                            & $\approx 2.4\times 10^7$                    \\
Full-physics, L62     & TNG-model             & $\Lambda$CDM, N5, N1     & 62                                    & 512$^3$               & 512$^3$               & 1.3$\times 10^8$                            & $\approx 2.4\times 10^7$                    \\
Full-physics, L25     & TNG-model             & $\Lambda$CDM, F6, F5     & 25                                    & 512$^3$               & 512$^3$               & 8.4$\times 10^6$                            & $\approx 2.2\times 10^6$                    \\
Full-physics, L25     & TNG-model             & $\Lambda$CDM, N5, N1     & 25                                    & 2$\times $512$^3$     & 2$\times $512$^3$     & 8.4$\times 10^6$                            & $\approx 1.6\times 10^6$                    \\
Non-rad               & Non-radiative         & $\Lambda$CDM, F6, F5     & 62                                    & 512$^3$               & 512$^3$               & 1.3$\times 10^8$                            & $\approx 3.6\times 10^7$                    \\
DM-only               & \multicolumn{1}{c}{-} & $\Lambda$CDM, F6, F5, F4 & 62                                    & 512$^3$               & 512$^3$               & 1.5$\times 10^8$                            & \multicolumn{1}{c}{-}                       \\
DM-only               & \multicolumn{1}{c}{-} & $\Lambda$CDM, N5, N1     & 62                                    & 512$^3$               & 512$^3$               & 1.5$\times 10^8$                            & \multicolumn{1}{c}{-}                       \\ \hline
\end{tabular}}
\caption{Box sizes and resolutions of the different sets from the $f(R)$- and nDGP-SHYBONE simulation. From left to right the columns show the simulation name suffix; hydrodynamical model used, cosmologies available in each run, comoving box size; number of dark matter particles; initial number of baryonic cells, dark matter particle mass; average baryonic particle mass.}
\label{tab:shybone-all}
\end{table*}

In evolved $f(R)$ universes, the fifth force has a significant effect on perturbations with scales smaller than the Compton wavelength, $\lambda_c$, where  
\begin{equation}
    \lambda_c = a^{-1}\left( 3\frac{ {\rm d}f_R}{{\rm d}R}  \right)^{\frac{1}{2}} ,
\end{equation}
where $a$ is the scale factor. 
For distances greater than $\lambda_c$ the force decays exponentially. 
This translates into an increased growth rate of cosmological linear density perturbations on scales smaller than $\lambda_c$.\\

In the simulations the model of $f(R)$-gravity proposed by \citet{Hu07} is adopted where $f(R)$ is assumed to have the form

\begin{equation}
    f(R) = -m^2 \frac{c_1 \left(-R/m^2 \right)^n}{c_2 \left(-R/m^2\right)^n + 1},
\end{equation}
where $m^2 \equiv 8\pi G \bar{\rho}_{\rm M,0}/3 = H_0^2 \Omega_{\rm M}$, $\bar{\rho}_{\rm M,0}$ is the background matter density at $z=0$, $H_0$ is the Hubble constant and $\Omega_{\rm M}$ the dimensionless matter density parameter at today. The parameter $n$ is set to $n=1$. The parameters $c_2$ and $c_3$ are selected in such a way that they fulfill the gravitational constraints measured in the solar neighborhood \citep{Will14}. Also, the model is able to reproduce the late-time expansion history of the Universe, with the appropriate selection of values for the parameters $c_1$ and $c_2$, as shown in \citet{Hu07}:
\begin{equation}
    \frac{c_1}{c_2} = 6 \frac{\Omega_\Lambda}{\Omega_m} ;
\end{equation}
and

\begin{equation}
    \frac{c_2 |R| }{m^2} \gg 1.
\end{equation}

With these considerations, it is possible to approximate the scalar degree of freedom, $f_R$, to:
\begin{equation}
    f_R \equiv \frac{{\rm d}f(R) }{{\rm d}R} = -n \frac{c_1 (R/m^2)^{n-1}}{[c_2 (R/m^2)^n + 1 ]^2} \approx -n\frac{c_1}{c_2} \left(\frac{m^2}{R}\right)^{n+1}.
\end{equation}

Finally, the scalar degree of freedom can be expressed in terms of the background value of the scalar field at $z=0$, $\bar{f}_{R0}$. This parameter sets the potential depth threshold at which the screening starts to be effective.

For this work, we consider two values of $\bar{f}_{R0}$, the F6 model, $\bar{f}_{R0} = -10^{-6}$, and the F5 model, $\bar{f}_{R0} = -10^{-5}$. The simulation suite also has a dark matter only run with an F4 model, $\bar{f}_{R0} = -10^{-4}$.  In the F6 model, screening starts at a relatively low gravitational potential depth and the model is in good agreement with most observational constraints \citep{Terukina14}. On the other hand, for F5 even regions with deep gravitational potentials can be unscreened and are in tension with the constraints presented in \citet{Will14}. Nevertheless, is still interesting to study the phenomenology of the model as it provides a useful tool to understand the properties of galaxies.

\subsubsection{The n-DGP model}
 
The Dvali-Gabadadze-Porrati braneworld model \citep{Dvali00}, assumes that matter in the Universe is confined to a 4-dimensional brane embedded in a 5-dimensional bulk space-time. The model presents a modification to the Einstein-Hilbert action, consisting of two arguments. The first is the classical Einstein-Hilbert action from General Relativity, and the second argument is the extension from the Einstein-Hilbert action to the 5-dimensions of the bulk as follows:
\begin{equation}
    S = \int_{brane} {\rm d}^4x\sqrt{-g}\left( \frac{R}{16\pi G}  \right) + \int {\rm d}^5x\sqrt{-g^{(5)}}\left( \frac{R^{(5)}}{16\pi G^{(5)}}  \right),
\label{eq:ndgp}
\end{equation}
where $g^{(5)}$, $R^{(5)}$ and $G^{(5)}$ correspond, respectively, to the equivalents of the determinant of the metric tensor, the Ricci scalar curvature and the gravitational constant in the space-time bulk.

From here, it is possible to define a characteristic length scale, $r_c$, at which the behaviour of gravity transitions from the 4-dimensional brane to the 5-dimensional bulk.
This scale is called the cross-over scale and is defined as follows:

\begin{equation}
    r_c = \frac{1}{2} \frac{G^{(5)}}{G}.
\end{equation}
 The change over the action produces modifications in the Friedmann equation, in the form of:
\begin{equation}
    \frac{H(a)}{H_0} = \sqrt{\Omega_{\rm M}a^{-3} + \Omega_{\rm DE}(a) + \Omega_{\rm rc}} \pm \sqrt{\Omega_{\rm rc}},
\end{equation}
from which, two branches of the DGP model come off; a self-accelerating one (sDGP) for which the positive value of $\sqrt{\Omega_{\rm rc}}$ is chosen, and the normal branch (nDGP) from which the negative value is chosen.
From now on we will only work with the normal branch, since it is able to reproduce the late-time cosmic acceleration without suffering from the ghost instabilities that exist in the self-accelerating branch.
$\Omega_m$ is the present-day value of the matter density parameter, and the $\Omega_{\rm DE}$ parameter is fixed in such a way that $H(a)$ matches that in a $\Lambda$CDM universe. Finally, $\Omega_{\rm rc}$ is defined as
\begin{equation}
    \Omega_{\rm rc} \equiv \frac{1}{4H_0^2r_c^2}.
\end{equation}

From these equations, we can see that the greater the value of $H_0r_c$, the more similar the model becomes to the standard $\Lambda$CDM model. In particular, for these simulations, values of $H_0r_c = 5$ and $H_0r_c = 1$ will be studied. These models will be referred to as N5 and N1 respectively.
These variations to the gravitational model lead to an enhancement in the gravitational potential of a factor of 1.12 for N1 and a factor of 1.04 for N5 at the present day.

\subsection{The SHYBONE Simulations}

The \textsc{shybone} simulation is the first suite of cosmological hydrodynamical simulations that simultaneously model galaxy formation, with a complete description of the subgrid physics, within modified gravity models. \citet{Arnold19} presented  \textsc{shybone-f(r)}, a simulation suite of model universes where gravity is described by the Hu \& Sawicki $f(R)$-gravity model. In \citet{Hernandez21} the second part of this suite introduced a set of universes with the nDGP gravity model.

\begin{figure*}
\centering
\includegraphics[width=\textwidth]{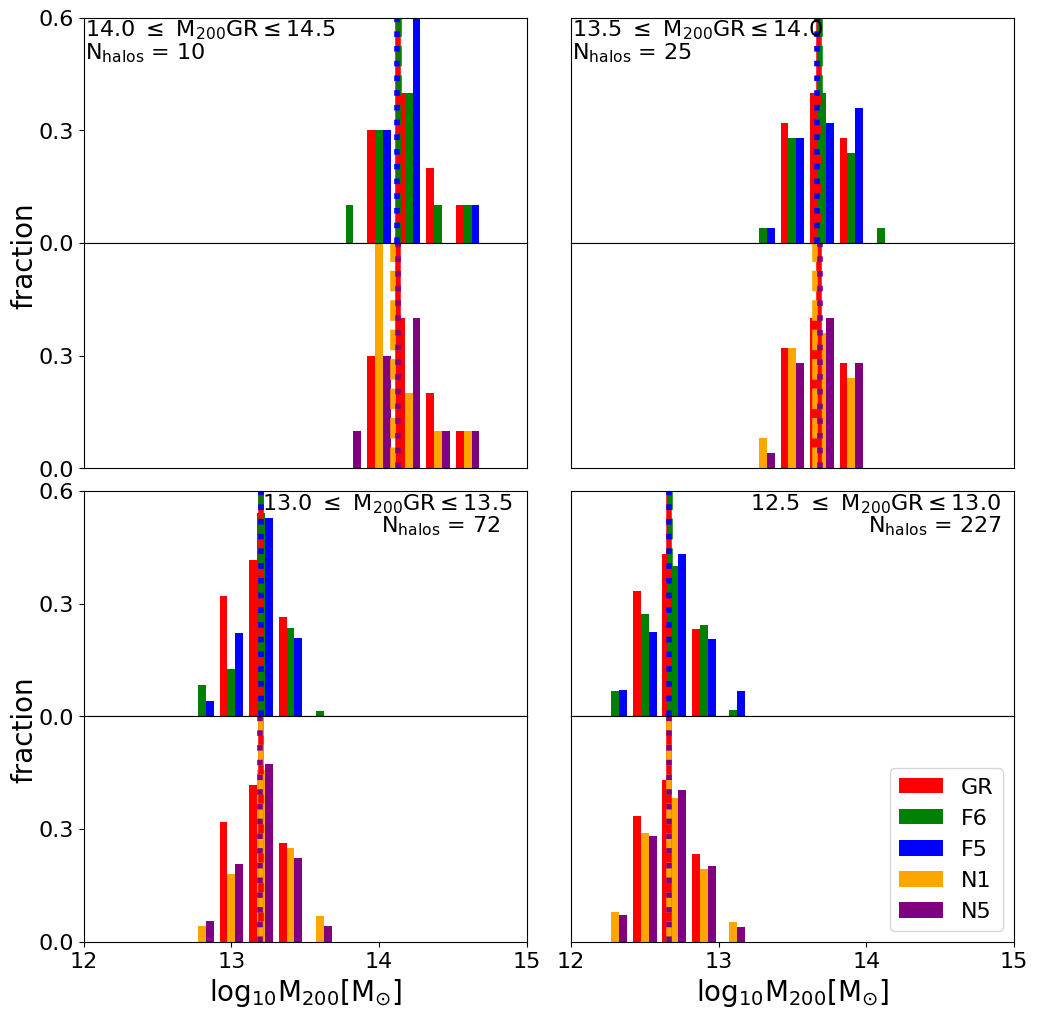}
\caption{Mass distribution of selected haloes in the MG models, obtained using the $M_{200}$ selection criterion (see text). Red, green, blue, orange and purple bars represent the GR, F6, F5, N1 and N5 host mass distribution of the selected haloes, respectively. Bins are 0.2dex wide for all models and are shown one next to the other for visualization purposes.
The corresponding $M_{200}$ mass ranges, as well as the number of selected haloes, are shown in the upper right corner of each panel. Dashed lines correspond to the median $M_{200}$ value obtained from the selected haloes in each model.}

\label{fig:dist_examples_m200}
\end{figure*}

\begin{figure*}
\centering
\includegraphics[width=\textwidth]{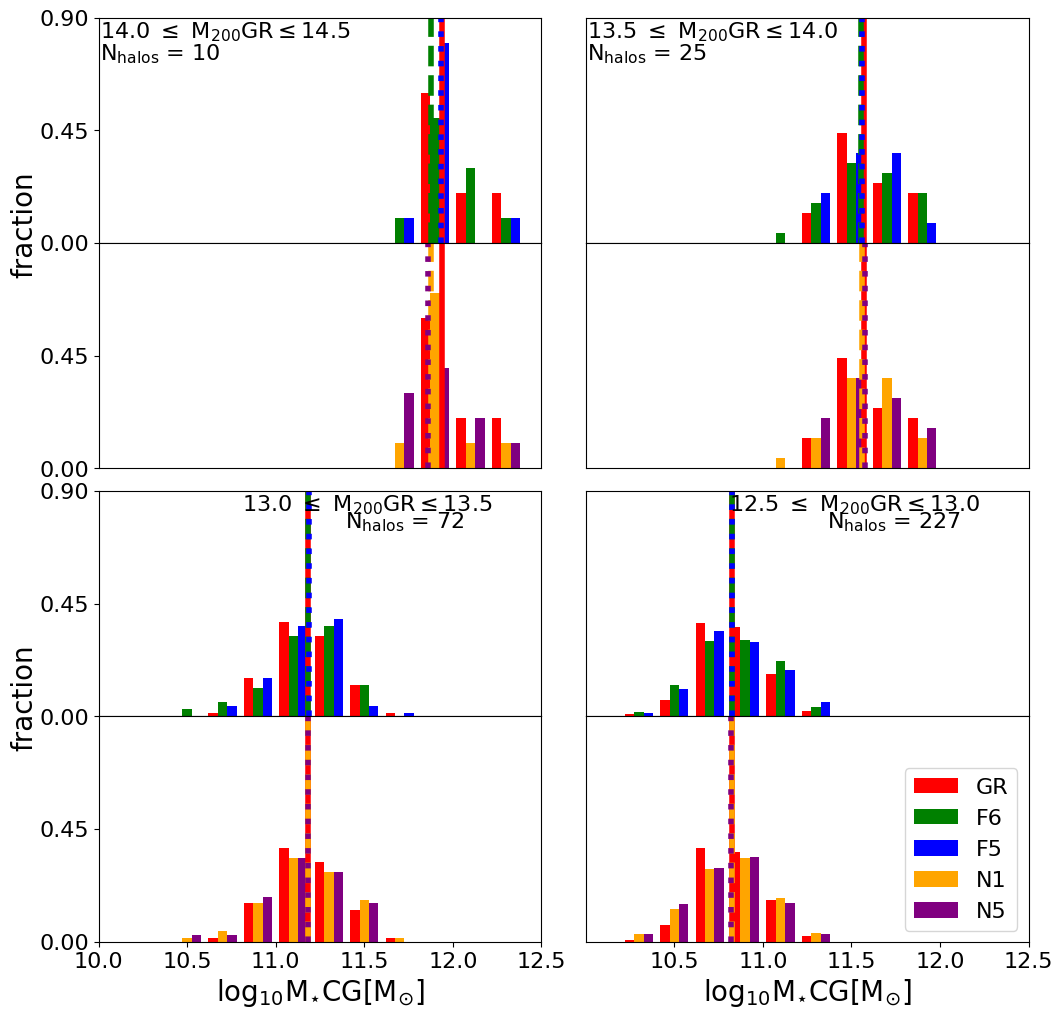}
\caption{As in Figure \ref{fig:dist_examples_m200}, for host halo distributions selected based on the CG $M_{\star}$ selection criterion (see text).}

\label{fig:dist_examples_bcgstar}
\end{figure*}

The simulations were performed using a modified version of the hydrodynamical simulation code \textsc{arepo} \citep{Springel10}, coupled with the Illustris-TNG galaxy formation model \citep{Pillepich18a}. To solve the equations of modified gravity presented in the previous section, a MG solver module was added to \textsc{arepo}.  The first module, specialized to solve the Hu \& Sawicki $f(R)$ model, was implemented by \citet{Arnold19}. The gravity solver for the nDGP model was added by \citet{Hernandez21}. These modules allow the equations for the scalar field and the Poisson equation to be solved in the quasi-static limit and are based on the modified gravity solver presented in the modified-gravity-\textsc{gadget} code \citep[\textsc{mg-gadget}, ][]{Puchwein13}. 
Some modifications were performed to the code, including using a more efficient method to solve the nonlinear field equations \citep{Bose17} and a local time-stepping scheme presented in \citet{Arnold16}.

The subgrid physics included in the Illustris-TNG galaxy formation model is based on the original Illustris galaxy formation model \citep{Vogelsberger14} and includes a set of well-calibrated prescriptions for the astrophysical processes needed to reproduce realistic galaxies in cosmological simulations. Among the processes included, in a subgrid fashion, there are prescriptions for black hole growth and AGN feedback, stellar feedback, galactic winds, gas cooling and UV-heating, an algorithm to compute the star formation rate and chemical enrichment. The parameters associated with the prescriptions mentioned above were fitted to allow the Illustris simulation to reproduce selected observational constraints considered as calibration datasets. These datasets are the galaxy stellar mass function at the present day, the gas fraction in galaxies, black hole masses and the cosmic star formation rate density. It should be noted that, for the modified gravity simulations, none of these parameters were changed from the original galaxy formation model (TNG). As shown in \citet{Arnold19} and \citet{Hernandez21},  the departure in the relations found in these simulations with respect to the observational data are smaller than the uncertainties in the observations.

The \textsc{shybone} simulation suite consists of 13 simulations for the the Hu \& Sawicki $f(R)$-gravity and 9 simulations for the nDGP model, corresponding to different choices for the MG parameters and resolution levels. A summary of the specifications for each $f(R)$ and nDGP run is presented in Table~\ref{tab:shybone-all}. All simulations were performed using cubic periodic boxes with periodic boundary conditions, sixteen with a box-size length $ L_{\rm box}$[$h^{-1}$Mpc]$ = 62$ and six with a box-size length $ L_{\rm box}$[$h^{-1}$Mpc]$ = 25$. The sixteen simulations performed in the large box share the same initial conditions, dark matter particle number ($N_{\rm DM} = 512^3$) and, for the hydrodynamical simulations, the initial number of gas cells. The large box subset is comprised of six simulations with the full-physics model for $\Lambda$CDM, F6, F5, N5 and N1 cosmology, three simulations with a basic, non-radiative hydrodynamic model for $\Lambda$CDM, F6 and F5 cosmology and seven dark matter only simulations for $\Lambda$CDM, F6, F5, F4, N5 and N1 cosmology. Additionally, six simulations in a smaller box are available for the full-physics model. These simulations were performed for $\Lambda$CDM, F6, F5, N5 and N1 cosmologies, and have roughly 15 times better resolution than their large-box counterpart. All simulations share the same cosmological parameters measured by the Planck mission \citep{Planck16}, with $n_s$ = 0.9667; $h \equiv H_0/100$ km s$^{-1}$Mpc$^{-1}$; $\Omega_{\Lambda} = 0.6911$; $\Omega_b = 0.0486$; $\Omega_{\rm m} = 0.3089$ and $\sigma_8 = 0.8159$, where $\Omega_{\rm m}$, $\Omega_{\Lambda}$ and $\Omega_{\rm b}$ correspond to the dark energy, baryonic density and matter densities respectively; $h$ is the normalized Hubble parameter; $\sigma_8$ is the square root of the linear variance of the matter distribution when smoothed with a top-hat filter of radius 8$h^{-1}$cMpc and $n_s$ is the scalar power-law index of the power spectrum of primordial adiabatic perturbations.
\begin{figure*}
\centering
\includegraphics[width=\textwidth]{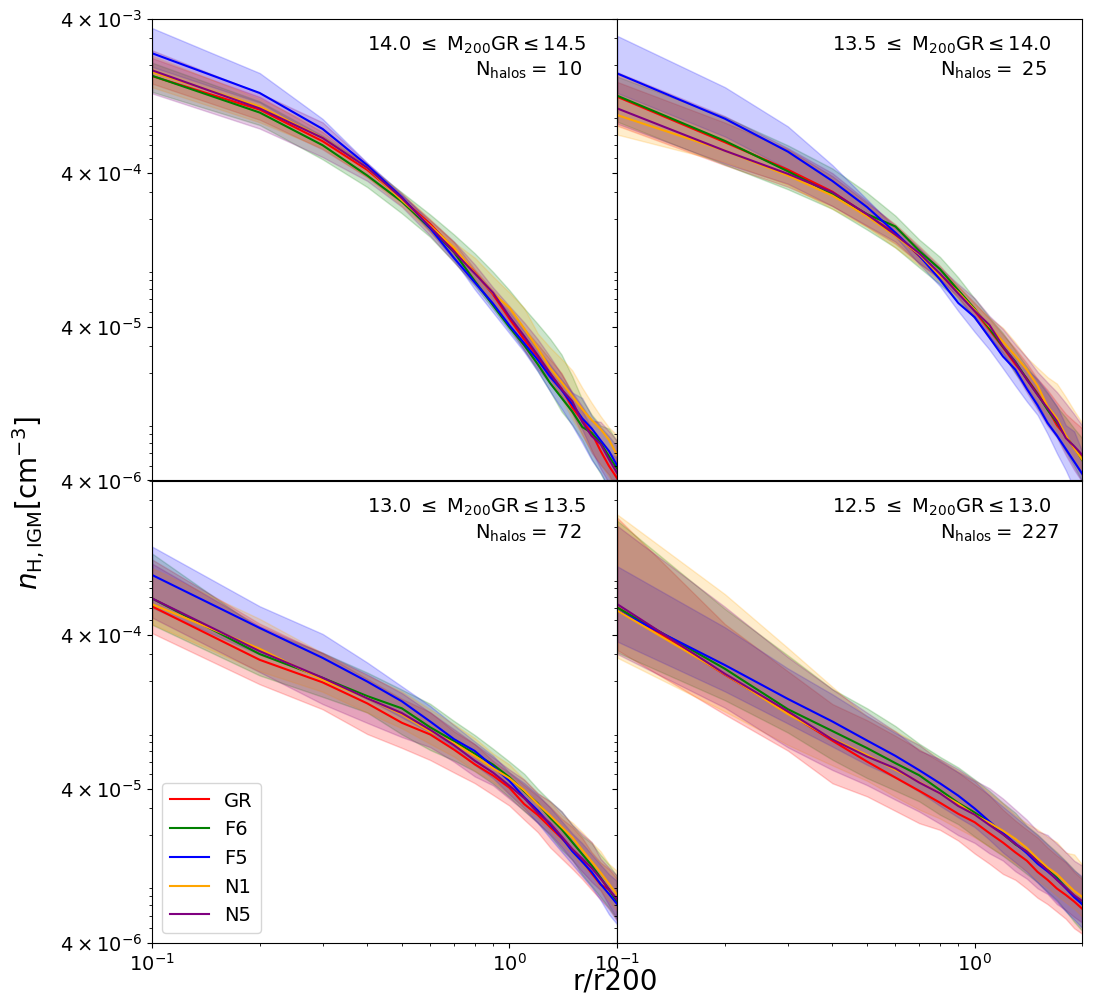}
\caption{Median of all azimuthally averaged gas density profiles for the selected haloes in the $f(R)$ and nDGP gravity models. Host haloes were selected based on the $M_{200}$ criterion. Red, green, blue, yellow and purple lines show the GR, F6, F5, N1 and N5 models respectively. The shaded areas indicate each distribution's 75 percentile central range, leaving 12.5$\%$ above and below. The corresponding $M_{200}$ mass ranges, as well as the number of selected haloes, are shown at the bottom of each panel.}
\label{fig:gasdistM_200}
\end{figure*}

\begin{figure*}
\centering
\includegraphics[width=\textwidth]{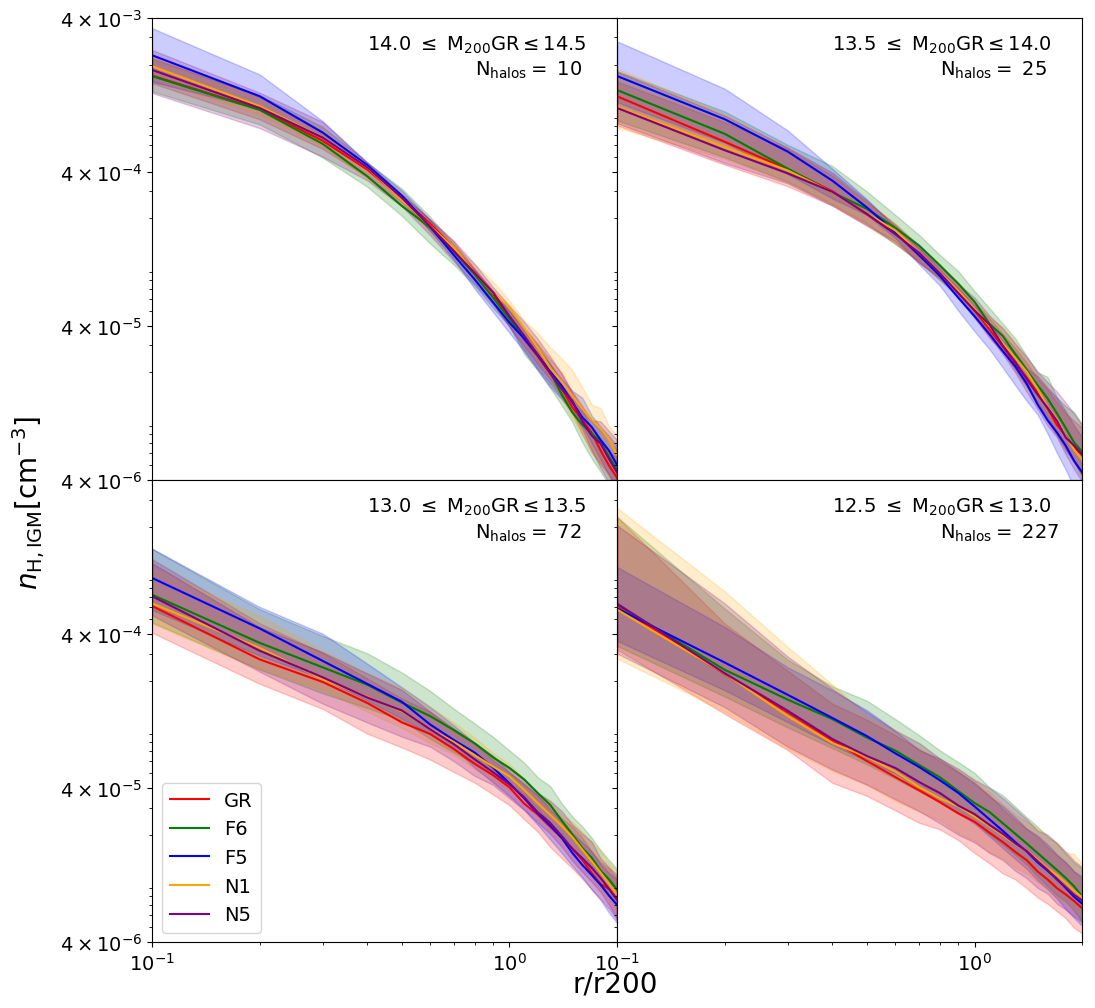}
\caption{As in Fig.~\ref{fig:gasdistM_200}, for host halo distributions selected based on the CG $M_{\star}$ selection criterion.}

\label{fig:gasdist_BCG}
\end{figure*}
The main products of the simulations were stored in 46 snapshots between $z=3$ and $z=0$ for the $f(R)$-gravity suite, and 99 snapshots between $z=20$ and $z=0$ for the nDGP-model. Based on these snapshots, a two-step procedure was performed to find the bound substructures using the \textsc{subfind} algorithm\citep{Springel01}. 

First, to define bound halos, a friends-of-friends (FoF) algorithm is applied to all dark matter particles using a linking length $b = 0.2$ times the mean interparticle distance. Baryons are then assigned to the FoF (if any) associated with their nearest dark matter particle. If a FoF halo possesses fewer than 32 dark matter particles, it is considered unresolved and discarded. As a second step, \textsc{subfind} identifies any gravitationally self-bound substructures (or `subhaloes') within a FoF halo taking dark matter and baryons into consideration. These subhaloes are identified as local overdensities using a binding energy criterion. For a more detailed description of the method, we refer to \cite{Springel01} and \cite{Dolag09}.
Here we will analyze the full-physics large box simulations of both suites to study the properties of galaxy populations inhabiting dense environments, considering different cosmologies. Our main goal is to understand the effect that different gravity models may have on the transformation from star-forming to quenched galaxies as a function of the environment in which they reside. In particular, we will focus on the intermediate-resolution large box simulations for the $f(R)$-gravity (GR, F6 and F5) and nDGP (GR, N5 and N1) runs. We define galaxies as all those subhaloes with a stellar content greater than $M_{\star} \geq 5\times10^9$M$_{\odot}$. As a result, we require at least a resolution of 100 star particles per object. For the halo selection, we use the $M_{200}$ given by the friends-of-friends algorithm to make the environmental separation and to assign cluster members.

\section{Comparing haloes from different models}
\label{sec:comp_haloes}

As discussed in the previous section, two haloes with the same mass but in different gravity models will produce different effective potentials acting over the rest of the structures. In the case of galaxy associations, this translates into different environmental effects exerted by, e.g., a galaxy cluster onto its members. As a result, the mass assembly of structures in different universes could follow different evolutionary paths, for both the baryonic and the dark matter components. Within this context, what we define as a galaxy cluster could differ from one gravitational model to another.

To address this potential problem, we used two observationally motivated criteria to compare haloes between different gravity models:

\begin{enumerate}
    \item To compare clusters based on their $M_{200}$.
    \item To compare clusters within a certain $M_{200}$ range based on the stellar mass of the central galaxy (CG). 
\end{enumerate}

The main advantage of using these criteria is that both can be applied to observational data by measuring properties that are not affected by the differences in the gravity models being compared here.
The first criterion may be derived observationally by using lensing estimations of the mass in groups and clusters, as neither $f(R)$ nor nDGP gravity affects the lensing potential, so this selection can be compared to what is done here for $M_{200}$.
On the other hand, the stellar mass responds to the total potential of the halo, given the physical processes that govern star formation such as gas cooling, as well as the feedback processes that regulate them depends directly on it and can be estimated from photometry.
Note that dynamical estimates of the mass are not the best option to compare between models, as these do depend on the modified potential.

By choosing these observationally measurable properties, we are able to make fair comparisons between models and to characterize differences between their galaxy populations. The procedure performed for the selection follows the same steps regardless of the property used for the comparison. First, we define four haloes $M_{200}$-mass bins in the GR-run to split haloes between galaxy clusters (log$_{10}M_{200}/$M$_\odot \geq 14$), high-mass groups ($13.5 \geq$ log$_{10}M_{200}/$M$_\odot \geq 14$), intermediate-mass groups ($13 \geq $log$_{10}M_{200}/$M$_\odot \geq 13.5$) and low mass groups ($12.5 \geq $log$_{10}M_{200}/$M$_\odot \geq 13$).
This sample is defined as the ``control sample'', and is used as a set of fiducial models. Second, for each MG model, we select sets of candidate haloes for comparison within the $M_{200}$ mass ranges previously discussed. Note that, to make sure sufficient candidates are selected, for the MG models the mass bins are enlarged by $\pm 0.3$dex. These candidates are subsequently sorted by mass. Starting from the lowest mass halo in each mass bin, we use a moving window to select a number of objects equal to the number of halos in the GR model. For each of these subsets the median of the desired quantity, i.e. $M_{200}$ or CG $M_{\star}$, is computed and compared to the corresponding value in the GR simulations. The subset with the closest median in each bin is used for the subsequent analysis.

Figs.~\ref{fig:dist_examples_m200} and \ref{fig:dist_examples_bcgstar} show the resulting distributions for selected haloes in each mass bin, using the $M_{200}$ and the CG $M_{\star}$  criteria, respectively. 
Each box represents a different environment. Boxes are divided into two panels, one for each MG model. The upper and bottom panels show the results for the $f(R)$-gravity models and the nDGP models, respectively. Red, green, blue, orange and purple bars stand for GR, F6, F5, N1 and N5 models, respectively.
We can see that, by the construction of our selection criteria, although the halo distribution is slightly different between models, the median value remains approximately the same at all mass bins. Small discrepancies in the median can be seen for the galaxy cluster mass bin, but this is expected given the small number of structures in this mass range available in the simulations. Nevertheless, the discrepancies between models median are $< 0.1$ dex.
A comparison between the samples obtained from both selection criteria shows that the set of MG cluster models selected based on $M_{\star}$, present slightly broader distributions than the corresponding GR distributions. The distribution of haloes is also broader when selecting them by $M_{\star}$ than when selecting them by $M_{200}$ in both, $f(R)$ and nDGP models.

\begin{figure*}
\centering
\includegraphics[width=0.7\textwidth]{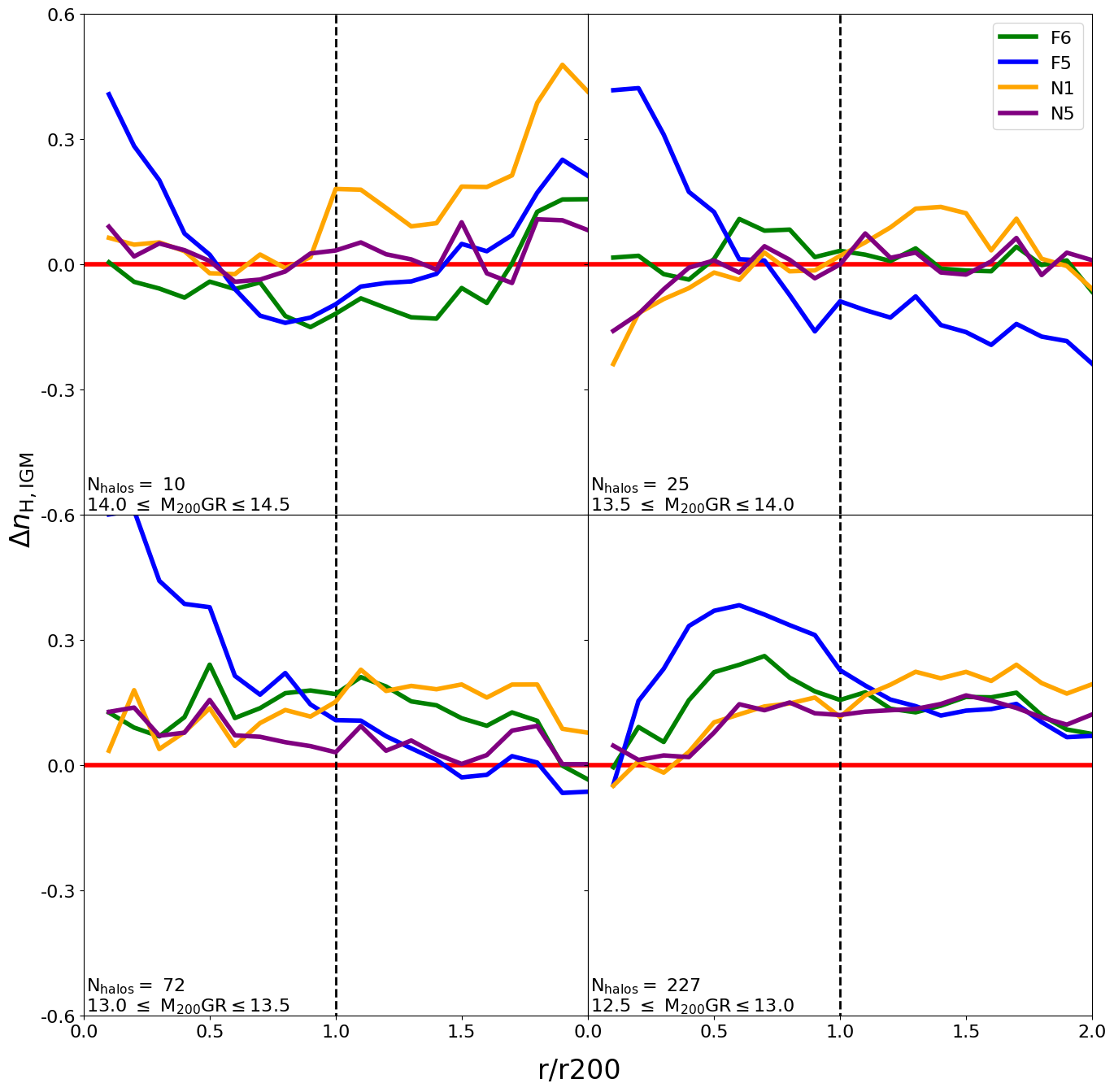}
\caption{Residual gas density profiles for MG haloes with respect to GR, when selecting by  M$_{200}$. Positive (negative) values denote an excess (decrement) in gas density. The corresponding $M_{200}$ mass ranges, as well as the number of selected haloes, are shown on the bottom of each panel. The red dashed line marks $r = r_{200}$. }

\label{fig:residual_M200}
\end{figure*}

\begin{figure*}
\centering
\includegraphics[width=0.7\textwidth]{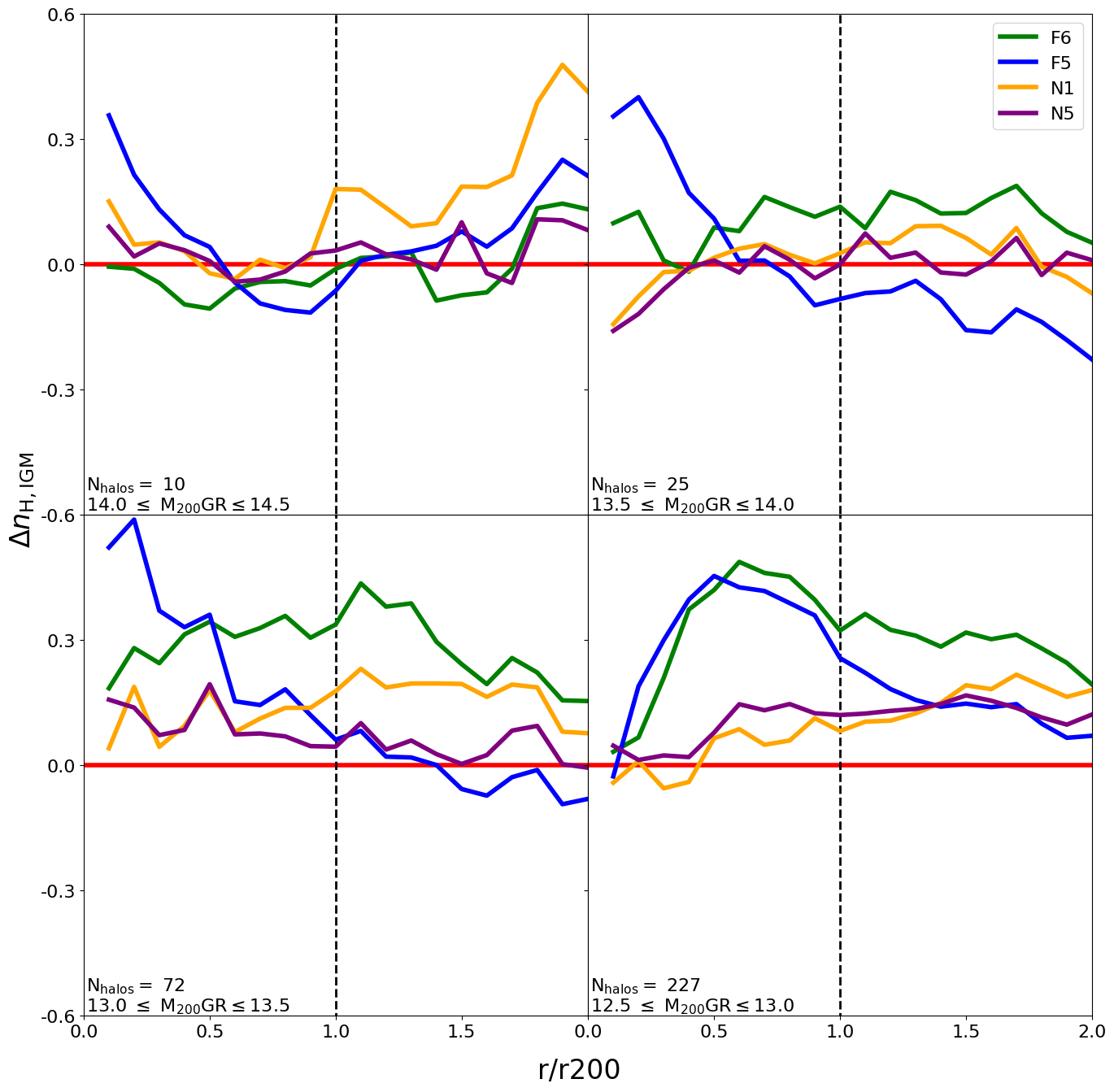}
\caption{As in Fig.~\ref{fig:residual_M200}, but for host halo distributions selected based on the CG $M_{\star}$ selection criterion.}

\label{fig:residual_BCG}
\end{figure*}

Our goal in this work is to characterize whether different gravity models leave different imprints on the observable properties of satellite galaxies. Thus, it is important to first explore whether the environment where these satellites evolve have different properties. In Figs.~\ref{fig:gasdistM_200} and \ref{fig:gasdist_BCG} we show the median azimuthally averaged gas density profile from the intra Group Medium (IGM), $n_{\rm H,IGM}$, for the selected haloes in each model for the most massive and least massive mass bin.
The left and right panels show the median gas density distribution for haloes in the $f(R)$ and nDGP gravity models respectively. The shaded area corresponds to 75$\%$ percentile for each model. Fig.~\ref{fig:gasdistM_200} focuses on the $M_{200}$ halo selection criterion. It is clear that when using  $M_{200}$ to select haloes, the GR and F6 models show little to no differences for values of $M_{200} \geq 10^{13}$M$_\odot$. However, the median haloes $n_{\rm H,IGM}$ profiles in the F5 model show significant discrepancies from GR in the inner regions  ($r < 0.5r_{200}$), especially for structures with $M^{\rm GR}_{200} \geq 10^{13}$M$_\odot$. For low mass groups,  no significant discrepancies between the F5, F6 and GR models are observed in the inner regions. However, it can be seen that, at $r \gtrsim 0.2 r_{200}$, the F5 and F6 gravity models present a slight overdensity with respect to GR. Such differences are hardly seen for the nDGP model, where the gas density distributions for both models, in all mass bins, are similar to the distributions obtained in GR. This is a consequence of the Vainshtein screening, which in nDGP models is very efficient in suppressing the fifth force inside haloes. This is unlike $f(R)$-gravity, where low-mass haloes can be fully unscreened and so, the effect of the fifth force may be felt even at the halo centre.

When using the CG $M_{\star}$ as the selection criterion, we find similar results. Discrepancies between the models are more evident for the $f(R)$ model, especially at intermediate masses. In particular, from Fig.~\ref{fig:gasdist_BCG} we can see that, while F5 shows differences with respect to GR in all mass bins, for $10^{14} \leq M^{\rm GR}_{200} \leq 10^{13.5}$M$_\odot$, the F6 model starts to differ from GR and becomes more similar to F5. For $M^{\rm GR}_{200} \leq 10^{13}$M$_\odot$, F5 and F6 show very similar behaviour, but both depart significantly from GR.

To better visualize the differences between the gas density distributions, in Figs.~\ref{fig:residual_M200} and \ref{fig:residual_BCG} we plot residual distributions, i.e. 

\begin{equation}
    \Delta n_{\rm H,IGM} = \dfrac{n^{\rm MG}_{\rm H,IGM} - n^{\rm GR}_{\rm H,IGM}}{n^{\rm GR}_{\rm H,IGM}}.
\end{equation}
Positive (negative) values represent regions that are overdense (underdense) with respect to GR. The black vertical dashed line indicates a clustercentric distance of $r = r_{200}$. 
In general, we find that in low-mass haloes (bottom panels) and for $r >0.3 r_{200}$, all MG models are denser than their GR counterparts, regardless of the selection criteria used. For massive groups and clusters, the distribution is much noisier due to the low number of halos. Thus, differences are less clear. As expected, the F5 model is the one that shows greater discrepancies with respect to GR in any mass bin.

This behaviour in $f(R)$-gravity haloes was previously reported by \citet{Mitchell19}. Low-mass haloes become unscreened earlier than larger haloes. Once that haloes become unscreened, the potential inside haloes depends by 4/3, so that gas is attracted towards the centre, leading to an enhanced profile, as can be seen for haloes with $M^{\rm GR}_{200} > 10^{13}$M$_{\odot}$.
On the other hand, in small haloes that have been unscreened for long enough, particle velocities have had enough time to also increase. This leads to an increase in kinetic energy, making the gas distribution manage to stay away from the centre regions, as can be seen for low-mass groups.
Finally, as mentioned before, for the nDGP model, the Vainsthein mechanisms efficiently suppress the fifth force, so little to no difference is expected. Above the virial radius, the screening becomes weaker, and the gas tends to be attracted towards the halo centre by the fifth force.

Even though the distributions do not significantly change by considering different selection criteria, selecting haloes by their CG $M_{\star}$ leads to slightly denser haloes in all MG models. This is especially clear for the lowest mass haloes in the $f(R)$-gravity. Regarding the nDGP model, the discrepancies with GR are relatively small compared to $f(R)$ haloes. The strongest differences between these models and GR are seen for the N1 models at the outskirts of galaxy clusters (top left panel).

\begin{figure*}
\centering
\includegraphics[width=\textwidth]{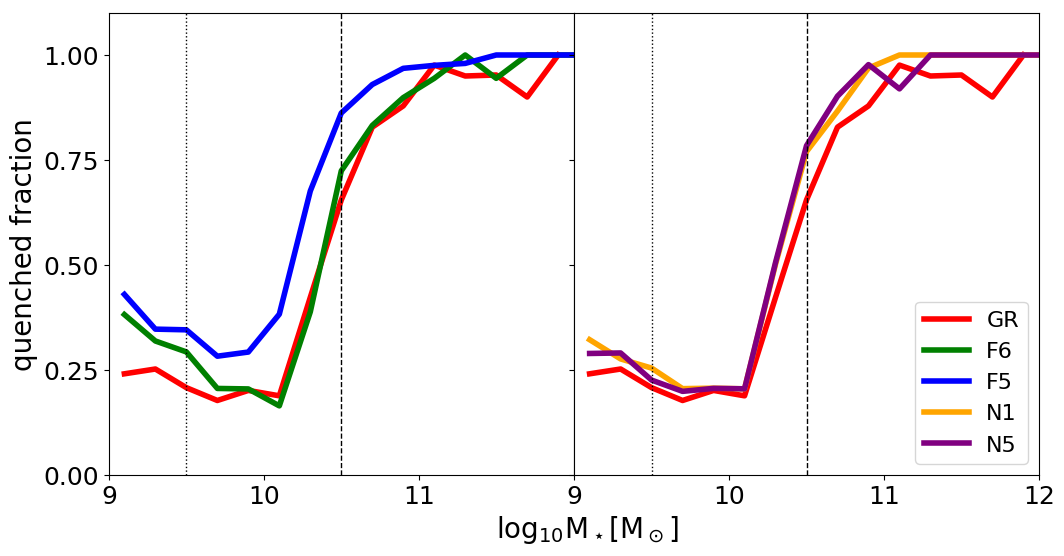}
\caption{Quenched fractions for all galaxies residing in our simulation boxes, independent of the environment in which they reside. Black dotted and dashed lines indicate the resolution threshold of $\sim 100$ and $\sim 1000$ stellar particles, respectively. The threshold in sSFR used to define galaxies as passive (quenched) is sSFR $< 10^{-11} yr^{-1}$. The left and right panels show the quenched distribution for galaxies in the $f(R)$ and the nDGP model, respectively.}

\label{fig:quenched_fractions_all}
\end{figure*}

Taking this into consideration, it can be expected that the galaxy population between models differs between them as well. In what follows, we will characterize the properties of their member galaxy population to understand the different imprints left by these discrepancies.

\begin{figure*}
\centering
\includegraphics[width=\textwidth]{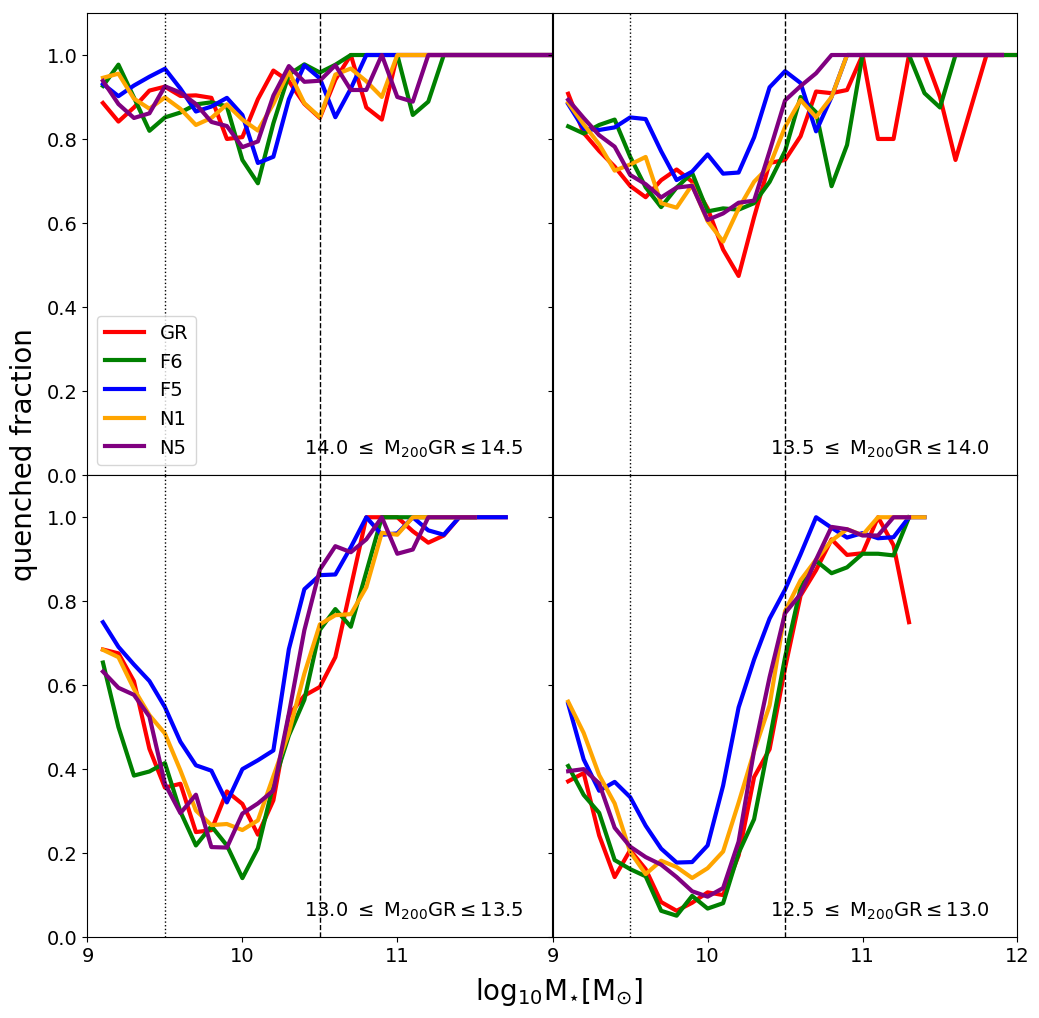}
\caption{Quenched fraction for galaxies as a function of their stellar mass and environments. Host haloes were selected using the $M_{200}$ selection criterion. Black dotted and dashed lines indicate the m $\sim 100$ and $\sim 1000$ stellar particles resolution thresholds, respectively. The threshold in sSFR used to define galaxies as passive (quenched) is sSFR $< 10^{-11} yr^{-1}$. Quenched fractions grow with stellar mass and towards denser environments. Differences between models become more evident at lower host halo masses ($M^{\rm GR}_{200} < 10^{13.5}$M$_{\odot}$). }

\label{fig:quenched_fractions_m200}
\end{figure*}

\begin{figure*}
\centering
\includegraphics[width=\textwidth]{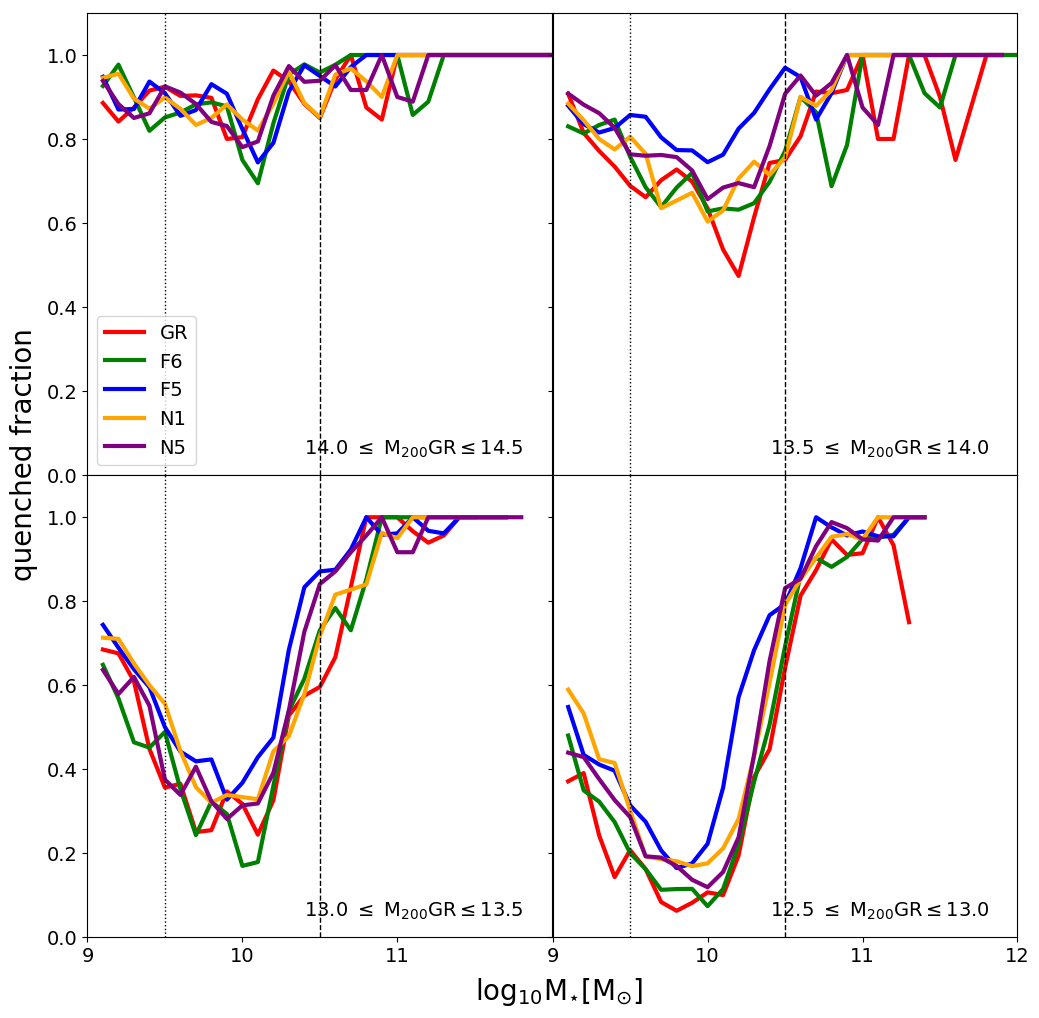}
\caption{As in Figure \ref{fig:residual_M200}, for host halo distributions selected based on the CG $M_{\star}$ selection criterion.}

\label{fig:quenched_fractions_bcg}
\end{figure*}

\section{Galaxy population in different models}

\label{sec:res}

All processes playing a role in galaxy formation and evolution are directly or indirectly linked to gravitational effects. As such, it is interesting to study how galaxies evolve in different gravity models. This is a topic that, to date, has received little attention due to the computational challenge of modelling galaxy formation in modified gravity cosmologies. In this section, we characterize some of the key properties that define populations of galaxies, such as passive fractions and colour distributions. 
Comparisons between models are made by using the selection criteria described in Section~\ref{sec:comp_haloes}.

\subsection{Quenched fraction}
\label{qf}
Understanding how galaxies become quenched can provide us important information about their different evolutionary pathways. Within this context, characterizing the fraction of quenched galaxies as a function of their environment can give us hints about how strongly 
galaxies are shaped by the environment in different gravity models.

In Fig.~\ref{fig:quenched_fractions_all} we first show the quenched galaxy fraction, considering all galaxies in the simulations, i.e., independent of the environment in which they reside. Quenched fractions are shown as a function of galaxy stellar mass, $M_*$. 

Following previous studies, to define a galaxy as quenched, we use a threshold in specific star formation rate (sSFR), defined as the instantaneous star formation rate (SFR) divided by the galaxy's total stellar mass, $M_*$. In particular, we define a galaxy as quenched if its sSFR $\leq$ 10$^{-11}$yr$^{-1}$ \citep[][]{Weinmann10,deLucia12,Wetzel12,Pallero19,Pallero20}.

Galaxies in the GR, F6, F5, N1 and N5 models are represented by red, green, blue, orange and purple lines, respectively in Fig.~\ref{fig:quenched_fractions_all}. This figure shows that, as expected, for $M_* > 10^{10} M_{\odot}$, the quenched fractions grow towards higher stellar masses, regardless of the model. However, MG models such as F5, N1 and N6 show larger quenched fractions than  GR. In particular, the quenched fraction in F5 is $\sim 20\%$ higher than that in GR at $M_* \lesssim 10^{11} M_{\odot}$. Conversely, F6 shows a similar quenched fraction distribution to the results found for GR. We note as well that the quenched fractions also start to rise for $M_* \lesssim 10^{9.5} M_{\odot}$. A similar result was already reported by \citet{Schaye15}, although with different simulations. \citet{Schaye15} shows that, for the \textsc{eagle} simulation, the quenched fraction starts to rise when galaxies fall below the $\sim 100$ particle stellar particles resolution limit, which is likely to be due to numerical noise effects.

The vertical dotted and dashed lines in Fig.~\ref{fig:quenched_fractions_all} indicate the $M_*$ values where galaxies contain 100 and 1000 stellar particles, respectively. It should be noted that these are present-day masses and that the number of dark matter particles per galaxy is typically a couple of order magnitudes greater than their number of stellar particles. Objects with stellar mass resolutions below that indicated by the dotted lines are discarded from our analysis. 
It is well reported in the standard model that the quenched fractions in galaxies increases toward higher stellar masses, due to inner mechanisms or, as is usually referred, due to  \textit{mass quenching}\citep[eg.][]{Peng10}. As all physical mechanisms are in one way or another related to the gravitational potential exerted by the galaxy itself, or the environment in which galaxies reside, changes in the quenched fraction are expected to happen between models, especially in those regions in which the screening mechanisms are less efficient.

As we are interested in characterizing our results as a function of environment, in 
Figs.~\ref{fig:quenched_fractions_m200} and \ref{fig:quenched_fractions_bcg} we show the quenched fractions of galaxies residing within groups and clusters in the five different gravity models.  The upper left panel shows the quenched fractions for galaxy clusters; the upper right panel shows massive groups, the bottom left shows results for intermediate-mass groups and the bottom right for low-mass groups. As before, we find that regardless of the model and the selection criteria,  for $M_{\star} \gtrsim 10^{10}$M$_{\odot}$ the quenched fraction grows towards higher stellar masses. In addition, and as expected, we find that this fraction also increases for more massive environments. In Fig.~\ref{fig:quenched_fractions_m200}, we show the results for all our models when host haloes are selected using the $M_{200}$ criterion. Of the 5 models presented, F5 is the one that typically shows the greater quenched fraction in any mass bin. This trend is more notable for the low and intermediate-mass groups (bottom panels). Two important things can be deduced from this result:

\begin{enumerate}
    \item Given the enhanced gravity experienced by member galaxies in regions where the fifth force is active, environmental effects start to gain relevance at lower halo masses.
    \item The enhanced gravity facilitates earlier gas consumption. This, for example, could be due to a starburst phase, galaxy mergers or early AGN activity. 
\end{enumerate}
  
These results will be explored further in future work, by following the evolutionary paths of individual galaxies in different models once the merger trees for the galaxies, in all simulations, become available.
  
Differences in the quenched galaxy fraction distributions for the other models N1, and N5 are not as clear as in Fig.~\ref{fig:quenched_fractions_all}. Nevertheless, these MG models present a higher quenched fraction than GR for intermediate and low-mass groups. This result is less evident for galaxy clusters. This is because, within these more massive galaxy structures, the effect of the environment is minor. As shown in \citet{Pallero19} and \citet{Pallero20}, most galaxies are quenched, regardless of their mass, when found within these large clusters. As before, the F6  model is the most similar to GR. 
  
In Fig.~\ref{fig:quenched_fractions_bcg} we show the distribution of the quenched galaxy fraction as a function of galaxy mass when environments are selected according to the CG $M_{\star}$. Similar to what was found for the density profiles of the host haloes, we found no significant differences when changing the selection criteria. The quenched fractions of galaxies follow the same trends shown in Fig.~\ref{fig:quenched_fractions_m200}, at any CG mass bin. As a result, the discrepancies observed in the quenched fraction can be associated with the different models, rather than the halo selection. These figures suggest that the criteria used for the host halo selection are more relevant for the structures themselves rather than for the galaxies residing within them, at least for galaxies at $z=0$.

In general, we can see that the quenched fractions in gravitational models where the fifth force has a larger radius of action are systematically higher with respect to GR, with F5 showing large discrepancies with GR, especially in the low-mass group bin.
One fascinating result that may be observed when splitting by the environment, is that even though N5 should be more similar to GR than N1, we can see large discrepancies with GR in the $10^{13} \leq M^{GR}_{200} \leq 10^{13.5}$[M$_{\odot}$] mass bin, similar to F5. 
This result suggests that the environment in the N5 model may start affecting the evolution of galaxies at less dense environments than in GR, given that when looking at the whole distribution of quenched galaxies, N1 and N5 show a rather similar distribution at any stellar mass. 
As the region where the Vainsthein screening is larger in the N1 model, galaxies evolving within this cosmology may be more resistant to starvation and/or ram-pressure than the N5 counterpart. In this sense, haloes of $10^{13}$M$_\odot$ may be massive enough to start stripping the gas from galaxies in N5. We acknowledge that to confirm this scenario, a more in-depth study should be carried out in which the gas depletion process of galaxies in different models is compared, and the ram-pressure and starvation process is characterized.
On the other hand, as expected, for models with gravitational potentials more similar to GR, the overall results in terms of quenched fractions are in much better agreement, regardless of the selection criteria.

\begin{figure*}
\centering
\includegraphics[width=\textwidth]{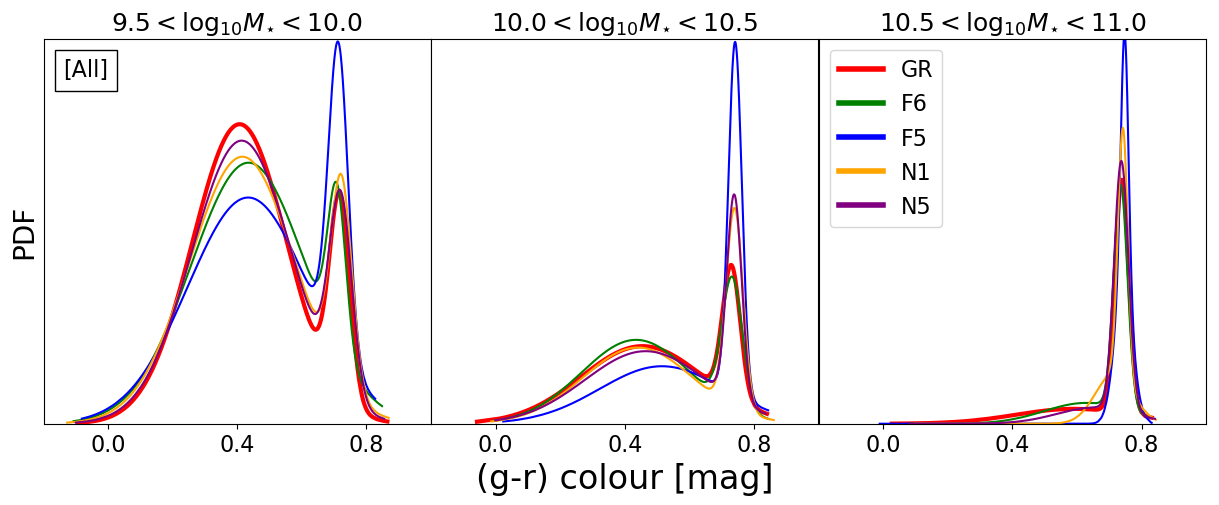}
\caption{Colour distributions for all galaxies residing in our simulation boxes, independent of the environment in which they reside. From left to right, the different panels focus on galaxies within different stellar mass ranges. To generate this figure, a double Gaussian distribution was fitted to each galaxy population, as described in the text. Red, green, blue, orange and purple lines represent galaxies belonging to GR, F6, F5, N1 and N5 models, respectively. Galaxies in MG models show a more predominant red population when compared to GR models. These trends are more evident when looking at lower stellar masses.}

\label{fig:colordist_stacked_m200}
\end{figure*}

\begin{figure*}
\centering
\includegraphics[width=\textwidth]{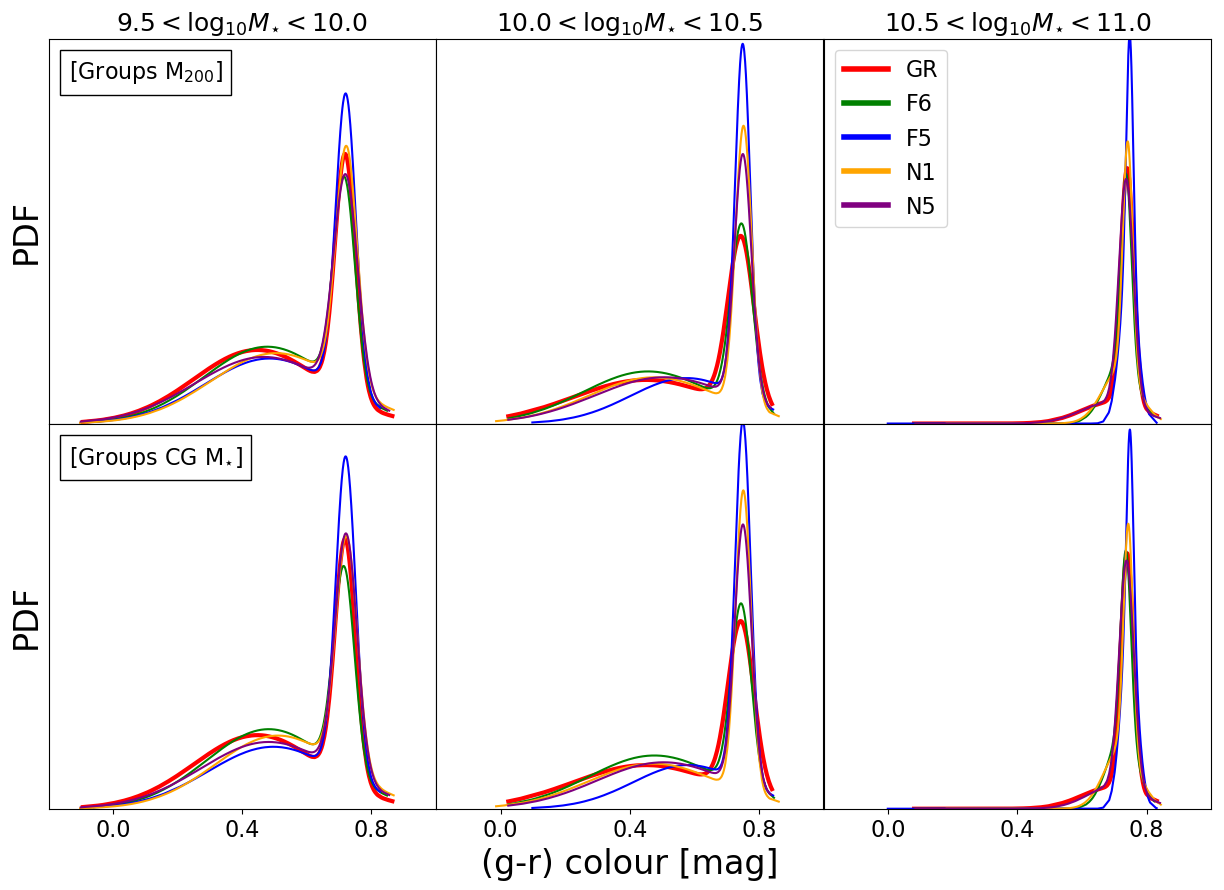}
\caption{The colour distribution of all galaxies found within halos more massive than $M^{GR}_{200} \geq 10^{12.5}$[M$_{\odot}$]. The top and bottom panels show the results obtained using the $M_{200}$ and the CG $M_{\star}$ selection criteria, respectively. 
Red, green, blue, orange and purple lines represent galaxies belonging to GR, F6, F5, N1 and N5 respectively. The only clear difference in this result when selecting by different criteria is that N5 is nearly identical to GR around the second peak for the CG M$_\star$ selection, but, along with F6, significantly lower than GR for the $M_{200}$ selection. For any other stellar mass bin, little to no differences can be seen when using different selection criteria to select haloes.
}

\label{fig:colordist_stacked_bcg}
\end{figure*}

\subsection{Colour distribution}

It has been widely known since the second half of the twentieth century that the colour of a galaxy reflects its predominant stellar population. Red colours are often associated with galaxies dominated by an old stellar population, and consequently with little to no recent star formation. On the other hand, blue galaxies reflect the presence of a large number of young stars and usually are currently forming stars.
Within this context, the distribution of galaxy colours in the Universe has proven to be strongly bimodal \citep{Strateva01,Baldry06,Manzoni2021}. Nevertheless, until now there has been no study exploring how different this distribution can be in universes evolved under different gravity models.

In the previous section, we showed that simulated galaxies in MG universes tend to show higher quenched fractions than the standard model. Following this, one might expect that MG galaxies should have redder colours. Nevertheless, as we go towards higher halo masses where the local environment dominates the quenching regime, we would expect these discrepancies to vanish. To explore this we follow a procedure similar to what was implemented in \citet{Baldry04} and further explored in \citet{Nelson18}; to isolate the red and blue galaxy populations. This is achieved by fitting a double Gaussian to the overall galaxy colour distribution. The following describes the procedure in detail:

\begin{enumerate}
    \item We select the population of galaxies residing within the desired environment.
    \item We split the corresponding galaxies by their stellar mass in three bins 0.5dex wide, over the interval $9.5<$log$_{10}M_{\star}/$M$_{\odot}<11$.
    \item For each stellar mass bin, we fit a double Gaussian to split between the red and blue galaxy populations.
\end{enumerate}

Following our approach in Section~\ref{qf}, we start by exploring the colour distribution of all galaxies in the simulations within ranges of stellar mass. That is, we consider not just galaxies in dense environments but galaxies from the field as well. The results are shown in  Fig.~\ref{fig:colordist_stacked_m200}. Red, green, blue, orange and purple lines represent the colour distributions obtained for the GR, F6, F5, N1 and N5 models respectively. From left to right, we show the colour distribution of galaxies separated in the three aforementioned stellar mass bins. It is clear that in the MG models galaxies show an overall redder distribution with respect to GR in all stellar mass bins. This is particularly clear for F5, N1 and N5 models, but less significant for F6. This is consistent with our previous results: i.e. denser IGM and larger quenched fractions for these models. 

\begin{figure*}
\centering
\includegraphics[width=\textwidth]{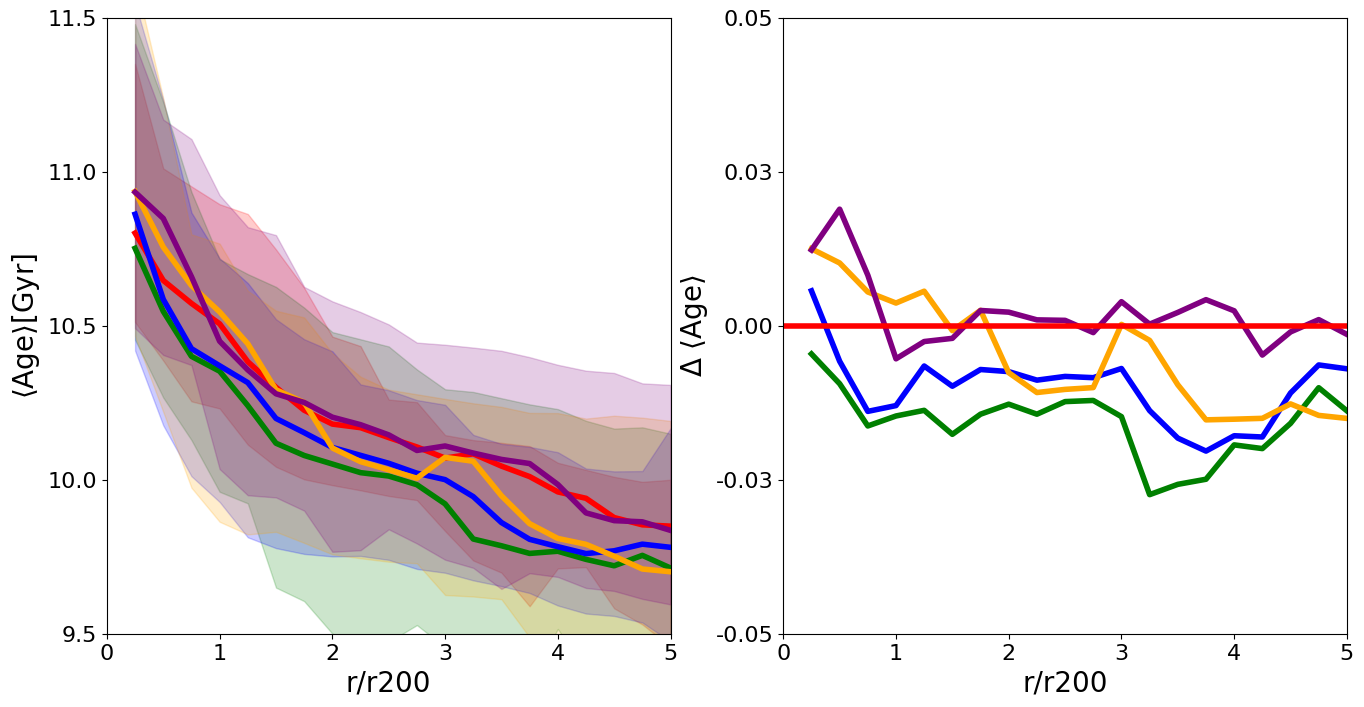}
\caption{\textit{left}: Mean age profile of the stellar population for galaxies surrounding clusters measured within one effective radii of the galaxy. The shaded areas represent the 25 to 75 percentile range for each distribution.
\textit{right:} Residual mean age profile for each modified gravity model, with respect to GR.
Red, green, blue, yellow and purple lines represent the average profiles for galaxies in the GR, F6, F5, N1 and N5 models respectively. Contrary to expectations, galaxies in the f(R) gravity model show, on average, younger stellar populations in the central parts of the galaxies regardless of the cluster-centric distance, when compared to GR. On the other hand, the nDGP model shows a significantly older stellar population at any cluster-centric distance with respect to GR. 
}

\label{fig:age_stacked_halfmassrad}
\end{figure*}

In Fig.~\ref{fig:colordist_stacked_bcg} we show our results for galaxies located within dense environments. As before, we select host halos based on their M$_{200}$ and CG $M_{\star}$. However, to increase the number statistics, we stack all satellites associated with dense environments (M$_{200} > 10^{12.5}$ M$_{\odot}$) into a single distribution. The top and bottom panels show the results when host halos are selected according to their M$_{200}$ and CG $M_{\star}$ values, respectively. In general, and as expected, in these dense environments, the distributions tend to show a more predominant red population with respect to what is shown in Fig.~\ref{fig:colordist_stacked_m200}, independent of the gravity model. This is due to the more significant role played by the environment in the evolution of satellite galaxies. As member galaxies enter groups and clusters, they rapidly get quenched \citep[e.g.][]{Pallero20}. Because most galaxies in groups and clusters are already quenched, it becomes more difficult to find strong differences between the MG models. A similar result can be seen in Figs.~\ref{fig:quenched_fractions_m200} and \ref{fig:quenched_fractions_bcg}. Nonetheless, the previous tendency towards a more predominant red population in the F5, N1 and N6 models with respect to GR is still present and clear.  As previously shown, F5 is the model that shows the largest discrepancies with respect to GR. Its overall population, even when looking at low stellar mass galaxies, is significantly more dominated by red galaxies. The bottom panel shows the same results for haloes selected according to their CG $M_{\star}$. The distribution is not affected by our selection criteria, indicating once again that our results are mainly associated with the different MG models.

\subsection{Age distribution}

As we mentioned before, the colour of a galaxy reflects its predominant stellar population. A more direct way to see this is by measuring the mean age of the stellar population. By comparing the mean stellar age of galaxies in different models, we can infer how the gravity model can affect the star formation history of galaxies residing in similar environments.
In particular, as we are interested in environmental effects, we will compare the age distribution for galaxies as a function of cluster-centric distance. 
To do this, we will measure the mass-weighted mean age for all the stellar particles within one-half mass radii, as follows:

\begin{equation}
    \langle \rm{Age} \rangle = \frac{\sum^{n}_{i = 1} m_i ~ age_i }{m_i}
\end{equation}
where $\rm m_i, \rm age_i$ corresponds to the mass and age of each star particle within our haloes.

We measure the mass-weighted mean age for all galaxies within 5$r_{200}$ of each galaxy cluster in all the models.
As we have 10 different clusters per gravitational model, as selected by our $M_{200}$ comparison criteria, to increase the signal and facilitate the comparison between models, we measured a median profile for each gravitational model by stacking galaxy ages in bins of cluster-centric distance as can be seen in the right panel of Fig.~\ref{fig:age_stacked_halfmassrad}. The shaded areas show the 25 to 75 percentile range for each distribution.

Here we can see that even though the quenched fractions are higher in the nDGP model in any bin of stellar or halo mass, the median ages in the N1 and N5 models are 2.5$\%$ ($\sim 300$Myr) older only closer to clusters (r$leq 1.5R_{200}$). After this threshold, the median stellar age of galaxies becomes the same as GR in the case of N5, and 2.5$\%$ ($\sim 300$Myr) younger for N1.

On the other hand, galaxies in the $f(R)-gravity$ model surprisingly show a slightly younger stellar population than their GR counterparts.
Even though the quenched fractions are higher and the colours are redder in the $f(R)$ model, galaxies in the $f(R)$ model have younger stars towards their centres.
This may be due to the presence of galaxies formed because of the modified gravity effect instead of primordial origin. The effect of the $f(R)$-gravity may drag down the average age of the stars in galaxies around clusters.
This result will be further explored in future work.

To highlight these results, similar to what was done in Section \ref{sec:comp_haloes}, in the right panel of Fig.~\ref{fig:age_stacked_halfmassrad}, we plot the residual distributions with respect to GR as follows:

\begin{equation}
    \Delta \langle \rm{Age} \rangle = \dfrac{\langle \rm{Age} \rangle^{\rm MG} - \langle \rm{Age} \rangle^{\rm GR}}{\langle \rm{Age} \rangle^{\rm GR}}.
\end{equation}
Positive (negative) values represent regions with older (younger) stellar population with respect to GR.
From this figure we can see that discrepancies between the nDGP model and GR grows towards larger clustercentric distances. For the $f(R)$ model, stars are mostly younger at any clustercentric distance, and the discrepancies remains mostly constant. The F6 model shows the youngest stellar age distribution with respect to any models.

\section{Summary and Future work}
\label{sec:summ}

Here, we presented the first steps towards a comprehensive study to characterize the impact that different gravitational models have on galaxy evolution. By using the state-of-the-art full-physics MG cosmological hydrodynamical simulations SHYBONE \cite{Arnold19,Hernandez21}, we measured the differences between the properties of galaxies residing in universes which adopt different gravity models. These discrepancies were characterized as a function of the environment in which galaxies reside, and their stellar mass.

In addition, to make the comparison between models as fair as possible, here we present two different selection criteria to compare subsets of haloes between simulations. One is based on the host $M_{200}$ and the second uses the central galaxy stellar mass (CG M$_{\star}$) within a certain range of $M_{200}$. 
These selection criteria become a powerful tool when compared with observations, as measurements of $M_{200}$ can be obtained through weak lensing and $M_\star$ can be directly obtained from photometry.
These two methodologies are independent of gravity, providing the same meaning in any model.

When looking at the median density profiles of the intra-group medium (IGM) we find that, in general, low- and intermediate-mass groups (M$_{200} < 10^{13}$ M$_{\odot}$) are typically denser than their GR counterparts, at any distance from their centres. This result suggests that groups and clusters undergo different assembly histories in the different gravity models and that this has a significant impact on the $z=0$ properties of their IGM. These results will be explored in more detail in a follow-up project. Results based on the different halo selection criteria show only marginally different results. For example, for the F6 model, when selecting haloes based on their $M_{200}$, the median gas density profile shows similar behaviour to the one displayed by GR, especially at high and intermediate host masses. However, when selecting haloes by their CG M$_{\star}$, the median F6 gas density profile shows slightly larger departures from the GR counterpart. 

Differences in the IGM properties could have an impact on the populations of galaxies residing within the corresponding environments. Our results also show that for those models with a more significantly enhanced gravity due to the action of a fifth force, the quenched fractions systematically grow and the galaxy populations become redder in general.  Models where the fifth force acts at larger scales (F5 and N5) are the ones that show the greater discrepancies with respect to GR, regardless of the host selection criteria. It is worth mentioning that, as shown in \citep[][]{Pallero20}, cluster members reach their quenching state within the first massive group they interact with and that the main culprit behind this process is a ram pressure stripping event. Thus, the observed differences in IGM density profiles for these MG models are expected to be behind this enhanced quenching process for the satellites. 

Contrary to what was expected, galaxies in $f(R)$ models, tend to be younger than in GR($\sim 300$Myr) near galaxy clusters, even though their population tends to be redder, and the quenched fractions tend to be larger. In the case of the nDGP model, galaxies show older ages near clusters ($\sim 300$Myr) as expected.

A key question to address in future work will be to test if previously known quenching mechanisms on GR, such as ram-pressure stripping, can affect galaxies evolving in MG models with similar efficiency. To more clearly characterize this effect, it is key to be able to follow the evolutionary history of individual galaxies. Understanding where and when galaxies suffer their transformation from star-forming to passive, their transition from the blue cloud to the red sequence, and the associated time scales will allow us to better constrain the differences between models. 

The results found in this project will provide important constraints on models from observations that will soon become available thanks to big galaxy surveys such as DESI \citep{DESI}, EUCLID \citep{EUCLID} and the Legacy Survey of Space and Time \citep{LSST}.



\section*{Acknowledgements}
DP acknowledges financial support from ANID through FONDECYT Postdoctrorado Project 3230379.
FAG acknowledges financial support from FONDECYT Regular 1211370. DP and FAG acknowledge funding from the Max Planck Society through a Partner Group grant. DP, FAG and YJ gratefully acknowledges support by the ANID BASAL project FB210003. We acknowledge financial support from the European Union’s Horizon 2020 Research and Innovation programme under the Marie Sklodowska-Curie grant agreement number734374-Projectacronym: LACEGAL.
NDP acknowledges support from a RAICES, a RAICES-Federal, and PICT-2021-I-A-00700 grants from the Ministerio de Ciencia, Tecnolog\'ia e Innovaci\'on, Argentina.
BL and CMB acknowledge support from the Science Technology Facilities Council through ST/X001075/1.  
This work used the DiRAC@Durham facility managed by the Institute for Computational Cosmology on behalf of the STFC DiRAC HPC Facility (\url{www.dirac.ac.uk}). The equipment was funded by BEIS capital funding via STFC capital grants ST/P002293/1, ST/R002371/1 and ST/S002502/1, Durham University and STFC operations grant ST/R000832/1. DiRAC is part of the National e-Infrastructure.

\section*{Data availability statement}

\bibliographystyle{mnras}
\bibliography{quench_final}




\bsp	
\label{lastpage}
\end{document}